\documentclass[twocolumn]{aastex62}
\usepackage{xcolor}
\usepackage{graphicx}

\usepackage{soul}

\graphicspath{{./}{figures/}}
\shorttitle{Detecting Gravitational Waves from Core Collapse Supernovae}
\shortauthors{Gill}
\begin{document}
\title{Milli-to-Deci-Hertz Detection Prospects for Gravitational Waves from Core-Collapse Supernovae}

\correspondingauthor{jasmine.gill@cfa.harvard.edu}

\author[0000-0003-4341-9824]{Kiranjyot Gill}
\affiliation{AstroAI; Center for Astrophysics \textbar{} Harvard \& Smithsonian, 60 Garden Street, Cambridge, MA 02138-1516, USA}

\begin{abstract} Gravitational wave (GW) astronomy truly began with the detection of merging compact binaries. The next breakthrough lies in detecting GWs from core-collapse supernovae (CCSNe), particularly the GW linear memory -- a phenomenon arising from aspherical matter ejection and anisotropic neutrino emission during stellar collapse. In this \textit{Letter}, we examine the feasibility of detecting this effect using next-generation space-based GW detectors or lunar-based GW observatories as this signature peaks below 25 Hz, which is largely inaccessible to terrestrial GW detectors due to seismic noise. Such a detection would provide fundamental insights into asymmetric matter dynamics near the collapsed core, shedding light on the stellar corpse that once represented the mass of the progenitor star and offering a front row seat to the potential formation of either a nascent neutron star progenitor or a black hole. Leveraging the longest-duration three-dimensional CCSNe simulations to date, spanning a wide range of progenitor masses, we show that space-based and lunar GW detectors present the most promising opportunities to extend the current CCSNe GW detection horizon ($\sim$ 10 kiloparsecs) to several megaparsecs. This may pave the way for regularly observing GWs from CCSNe beyond our Milky Way by utilizing the distinctive environments of space and the Moon to augment terrestrial detection efforts all the while synergistically advancing sophisticated data analysis techniques for routine GW detection.
\end{abstract}
\section{Introduction} \label{sec:intro}

Core-collapse supernovae (CCSNe), the cataclysmic deaths of massive stars ($\geq 8$ M$_\odot$), represent both a compelling and challenging detection prospect in gravitational wave (GW) astronomy. The field itself was inaugurated by the detection of a binary black hole (BBH) merger, GW150914 \citep{2016PhRvL.116f1102A}, and soon blossomed into multi-messenger astrophysics with the detection of the binary neutron star (BNS) merger, GW170817 \citep{2017PhRvL.119p1101A}. However, detecting GWs from CCSNe poses significant challenges due to their inherently lower GW energy emissions compared to compact binary mergers, which is due to the frequency dependence of GW energy, $E_{\rm GW} \sim f_{\rm GW}^2$. For CCSNe, $E_{\rm GW}$ ranges from $10^{-11}$ to $10^{-8}$ M$_\odot \mathrm{c}^2$, while compact binary mergers such as GW150914 radiate \( \sim 3 \, \mathrm{M}_\odot \mathrm{c}^2 \). The energy emitted in GWs from CCSNe is also highly dependent on progenitor characteristics, such as mass, rotation, and explosion/implosion asymmetry (see \citealt{2023PhRvD.107j3015V} and references therein).

The stochastic nature of the GW signals from CCSNe, driven by variations in progenitor dynamics and temporal evolution, further complicates their detection. Three-dimensional simulations have revealed distinct GW emission phases during stellar collapse and core bounce at supranuclear densities (see \citealt{2020LRCA....6....3M, 2021Natur.589...29B, 2023IAUS..362..215M, 2024arXiv241012105P} for recent reviews). These phases typically begin with an initial burst of emission, triggered by prompt convection within $\sim$ 0.3 seconds after the core bounce, followed by intermediate emission driven by turbulent fluid dynamics within the first second. Beyond this, GW emission persists, sustained by turbulence and proto-neutron star (PNS) oscillations. High-frequency emissions (100–2000 Hz) during the first second are primarily linked to violent motions of turbulent matter and oscillatory modes of the PNS. 

Sub-Hz GW emissions provide invaluable insight into the explosion dynamics of CCSNe. These signals are primarily generated by anisotropic neutrino emissions \citep{1978ApJ...223.1037E, 1996PhRvL..76..352B, Kotake_2004, 2012A&A...537A..63M,  takiwaki_18, 2021arXiv210505862M, 2022PhRvD.106d3020M, 2022PhRvD.105j3008R} and large-scale asymmetric convective motions of ejected matter \citep{Morozova_2018, vartanyan_19, Radice_2019, 2020ApJ...901..108V, 2023PhRvD.107j3015V}. A hallmark emission effect arises in the form of \textit{GW linear memory}, a persistent deformation in the space-time metric caused by asymmetric neutrino flux and anisotropic mass ejection, leaving a lasting imprint that does not average out over time. This lasting signal continues beyond the cessation of neutrino emissions after shock breakout, preserving critical information about explosion asymmetries and compact remnant formation \citep{1972ARA&A..10..335P, 1974SvA....18...17Z, 1978ApJ...223.1037E, 1987Natur.327..123B, Christodoulou:1991cr, PhysRevD.45.520, 1995ApJ...450..830B, 1997A&A...317..140M, vartanyan_19, 2020ApJ...901..108V, 2023PhRvD.107j3015V, 2024MNRAS.531.3732S, 2024MNRAS.532.4326P, 2024arXiv240801525C}. The amplitudes of these low-frequency emissions can reach 100 to 1000 cm, significantly larger than the amplitudes at the $\sim$ few centimeter scale of high-frequency emissions sourced from motion of matter alone \citep{vartanyan_19, 2020ApJ...901..108V, 2024MNRAS.531.3732S}.

Detecting low-frequency GWs, including the GW linear memory effect, presents significant challenges. Terrestrial detectors such as LIGO and Virgo \citep{2015CQGra..32b4001A, 2020JCAP...03..050M, Brady2020}, even with planned upgrades such as A$\#$ \citep{ligoLIGOG2301738v3LIGO, 2023arXiv230710421G}, Cosmic Explorer (CE) \citep{2020JCAP...03..050M}, and Einstein Telescope (ET) \citep{2019BAAS...51g..35R, 2020JCAP...03..050M, Branchesi_2023}, are fundamentally constrained by seismic noise below 25 Hz \citep{1984PhRvD..30..732S, 2023arXiv230710421G, ligoLIGOG2301738v3LIGO}. This limitation hinders their ability to observe the wealth of information contained within low-frequency GW signals from CCSNe. 

Furthermore, the sensitivity of terrestrial detectors at high frequencies remains insufficient to detect CCSNe signals beyond the Milky Way, restricting their reach to Galactic events within $\sim$10 kpc \citep{2016PhRvD..93d2002G, 2021PhRvD.104j2002S}. Slowly rotating, massive progenitors expected to dominate the local universe within 20 Mpc \citep{1972ARA&A..10..335P, 2005ApJ...626..350H} are therefore likely to remain undetected without transformative advancements in detector design. Although proposed facilities like CE and ET promise substantial improvements in sensitivity, the constraints imposed by Earth’s seismic environment underscore the need for alternative approaches to access this critical low-frequency regime.

Space-based GW detectors offer a promising solution to the challenges posed by terrestrial seismic noise. Although the Laser Interferometer Space Antenna (LISA) \citep{2017arXiv170200786A} is not optimized to detect GW from CCSNe, future missions such as DECIGO \citep{2021PTEP.2021eA105K} could extend the detection horizon to megaparsec distances \citep{2024arXiv240801525C, 2024arXiv241012105P}. Alternatively, deploying GW detectors on the Moon provides an innovative approach, leveraging the Moon's exceptional seismic isolation to achieve sub-Hz sensitivity \citep{Stebbins1990, 2024arXiv240207571C, Harms_2021, Jani_2021}. The natural quiet environment of the Moon and the low temperatures in permanently shadowed craters make it an ideal location for sustained GW observations \citep{2020CQGra..37u5011A, 2023AcAau.211..616R, 2023SSRv..219...67B, 2024arXiv240409181A}.

The Lunar Gravitational Wave Antenna (LGWA) \citep{Harms_2021, 2024arXiv240409181A} is designed to use inertial sensors to detect vibrations on the lunar surface caused by GWs. In contrast, the Laser Interferometer Lunar Antenna (LILA; \citealt{LILACollaboration}) would operate in a triangular configuration with arms spanning tens of kilometers. Both concepts are proposed for locations near the Moon's poles, such as the Shackleton, Shoemaker, or Oppenheimer craters \citep{MITUSOV2023105795, BASILEVSKY2024105839}. Although still in the development stage, LGWA and LILA hold significant promise for deployment within the next two decades, supported by initiatives such as NASA's Artemis program \citep{nasaArtemisDeep} and the European Space Agency’s lunar gateway program \citep{ESA2019MoonStrategy}.

In this \textit{Letter}, we explore how neutrino-driven GW emission models from CCSNe can extend detection horizons from 10 kpc to megaparsec scales using future space-based and lunar GW observatories. In Section \S\ref{sec:meth}, we present GW strain predictions from state-of-the-art three-dimensional CCSNe simulations, emphasizing the GW linear memory effect. Section \S\ref{sec:res} evaluates the detectability of these strains at 10 kpc by comparing them against sensitivity curves from a wide array of current and proposed detectors, including terrestrial observatories like LIGO O3, A$\#$, Cosmic Explorer (CE), and the Einstein Telescope (ET); deci-Hz space missions such as B-DECIGO, DECIGO, the Big Bang Observatory (BBO) \citep{2006PhRvD..73d2001C}, the Deci-Hz Observatory (DO) \citep{2020CQGra..37u5011A, 2021hgwa.bookE..50I}, and the Advanced Laser Interferometer Antenna (ALIA) \citep{2020CQGra..37u5011A, 2021hgwa.bookE..50I} alongside lunar detectors like the Gravitational-wave Lunar Observatory for Cosmology (GLOC) \citep{Jani_2021}, the Lunar Gravitational-Wave Antenna (LGWA), and the Laser Interferometer Lunar Antenna (LILA). In Section \S\ref{sec:con}, we discuss the broader implications of these findings for the future of GW astronomy and summarize our conclusions.

\begin{figure*}[h!] 
\centering 
\includegraphics[width=\textwidth]{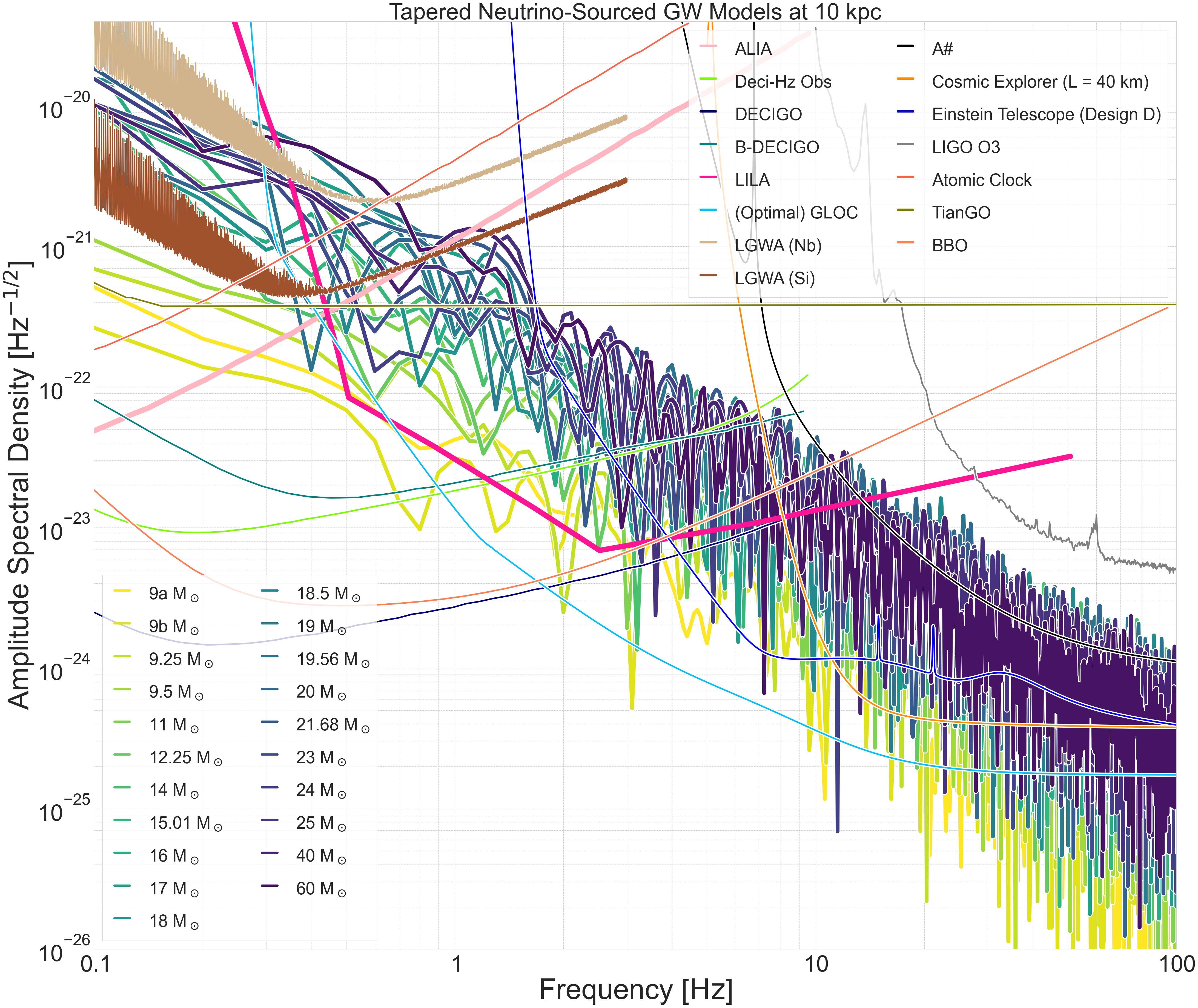}
\caption{Amplitude spectral densities of GW signals from neutrino-sourced emission models in CCSNe are shown for a source at 10 kpc, compared to the noise curves \(\sqrt{S_n}\) of various current and proposed GW detectors. These include lunar-based detectors (LGWA, GLOC, and LILA), deci-hertz observatories (B-DECIGO, DECIGO, and Deci-Hz Obs), space-based detectors (TianGO, Atomic Clock, BBO), and terrestrial observatories (LIGO O3b, A$\#$, Cosmic Explorer, and the Einstein Telescope). The figure demonstrates that GW signals from neutrino memory effects are significantly stronger than those from matter-sourced emission, particularly at low frequencies (below 1 Hz). This enhanced signal strength arises from the anisotropic emission of neutrinos during the collapse phase, which produces a long-duration, nearly constant offset in the GW strain. These signals are best captured by low-frequency detectors such as lunar-based observatories and deci-hertz detectors, which are uniquely sensitive to the low-frequency memory signature. Both axes use logarithmic scaling to highlight the wide frequency range and amplitude variations}
\label{fig:neutrino} 
\end{figure*} 

\begin{figure*}[h!] 
\centering 
\includegraphics[width=\textwidth]{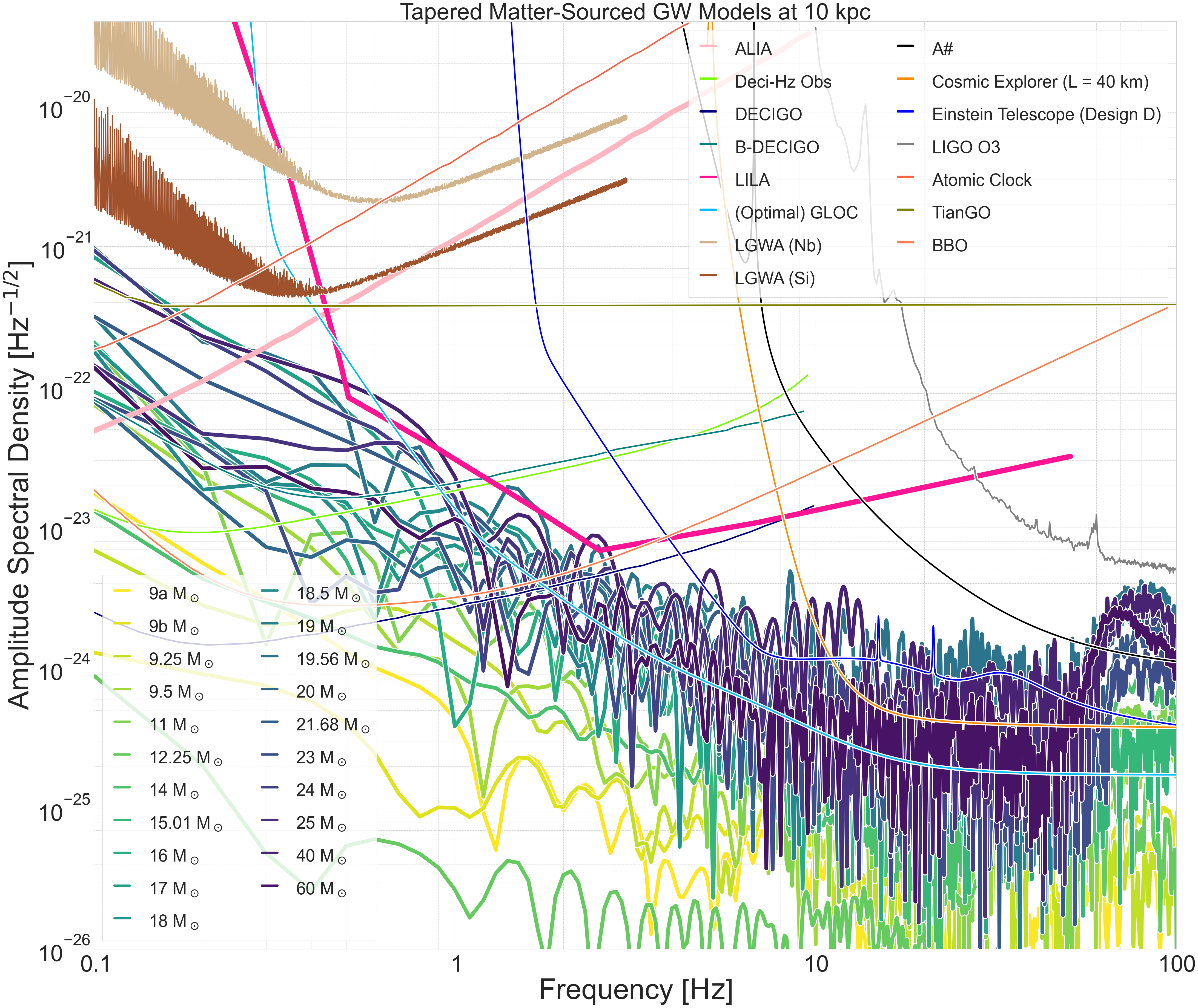}
\caption{Amplitude spectral densities of GW signals from matter-sourced emission models in CCSNe are shown for a source at 10 kpc, compared to the noise curves \(\sqrt{S_n}\) of various current and proposed GW detectors. These include lunar-based detectors (LGWA, GLOC, and LILA), deci-hertz observatories (B-DECIGO, DECIGO, and Deci-Hz Obs), space-based detectors (TianGO, Atomic Clock, BBO), and terrestrial observatories (LIGO O3b, A$\#$, Cosmic Explorer, and the Einstein Telescope). The analysis spans progenitor masses from 9 M$_\odot$ to 60 M$_\odot$, revealing trends in GW amplitude with increasing mass. Signals below 100 Hz probe large-scale mass motions during the collapse phase, best observed by lunar and deci-hertz detectors, while signals above 100 Hz reveal proto-neutron star formation and dynamics, accessible to terrestrial detectors. The figure highlights the unique sensitivity of lunar-based detectors in the low-frequency regime, emphasizing their potential to uncover features otherwise undetectable with terrestrial and space-based observatories. Both axes use logarithmic scaling to illustrate the broad frequency range and amplitude variations.}
\label{fig:matter} 
\end{figure*}

\section{Core-Collapse Supernovae Gravitational Wave Models} \label{sec:meth}

The non-rotating\footnote{$\Omega_0$ = 0 rad/s} GW waveform models from \citealt{2023PhRvD.107j3015V} and \citealt{2024arXiv240801525C} span progenitor masses from 9 to 60 M$_\odot$, as summarized in Table~\ref{tab:models}. These simulations represent the longest-running three-dimensional CCSNe waveform models to date, capturing detailed fluid instabilities and asymmetric mass ejections that drive GW emission across a broad range of frequencies. For further details on these models, we refer the reader to \citealt{2023PhRvD.107j3015V} and \citealt{2024arXiv240801525C}.

Lower-mass progenitors, such as the 9a and 9b models, provide the necessary insight into how GWs respond to the initial conditions of stellar collapse. The 9a model includes initial perturbations to the stellar structure, producing vigorous explosion dynamics, whereas the 9b model omits these perturbations, serving as a baseline for comparison. The interaction between early prompt convection and accretion plumes near the PNS results in GW amplitudes varying by up to three orders of magnitude, depending on the collapse dynamics tied to progenitor mass.

The evolutionary fate of each progenitor is strongly tied to its mass.~Most models lead to credible PNS formation, with GW emissions reflecting the dynamic nature of the explosion. Certain progenitors, such as those with masses of 12.25 M$_\odot$ and 14 M$_\odot$, fail to explode and instead collapse directly into stellar-mass black holes (BHs) within two seconds. These events produce brief GW signals associated with the initial collapse, accompanied by negligible GW energy from outward mass ejection. In contrast, higher-mass progenitors, such as the 19.56 M$_\odot$ and 40 M$_\odot$ models, initially explode, but later collapse into heavier stellar-mass BHs. These collapses generate distinct GW signatures, beginning with an initial surge in energy during the explosion phase, followed by a rapid decline as the BH forms.~Such stellar-mass BHs, formed from failed explosions or fallback processes, contribute to the population of compact remnants observed in the universe and align with the diverse stellar-mass black hole populations detected by the LIGO-Virgo-KAGRA collaboration.

The GW linear memory effect is a key feature of these models, arising from anisotropic neutrino emissions and asymmetric mass ejection during the explosion. This sustained effect appears across all progenitor masses, with detectability varying based on the progenitor's mass and explosion asymmetry. Lower-mass progenitors, such as the 11 M$_\odot$ and 23 M$_\odot$ models, exhibit enhanced GW strains that peak around 25 Hz due to pronounced asymmetries. Regardless of the final outcome—successful explosion or BH formation—the GW linear memory effect persists beyond truncation, emphasizing the importance of neutrino anisotropy in shaping the long-term GW signal profile.

\begin{deluxetable}{lccc}[t!]
\tablecolumns{4}
\tablecaption{Gravitational Wave (GW) Waveform Models from Core-Collapse Supernovae \label{tab:models}}
\tablehead{
\colhead{Progenitor Mass} & 
\colhead{Duration} & 
\colhead{$\mathrm{E_{\rm GW}}$} &
\colhead{Compact} \\
\colhead{(M$_\odot$)} & 
\colhead{(s)} &
\colhead{($10^{-10}$ M$_\odot {\rm c^2}$)} &
\colhead{Object}
}
\startdata
9a & 1.76 & 0.26 & NS \\
9b & 2.00 & 0.35 & NS \\
9.25 & 3.48 & 2.93 & NS \\
9.50 & 2.13 & 5.56 & NS \\
11 & 3.07 & 53.10 & NS \\
12.25 & 2.05 & 36.70 & \textbf{BH*} \\
14 & 2.70 & 69.70 & \textbf{BH*} \\
15.01 & 4.37 & 128.58 & NS \\
16 & 4.15 & 35.70 & NS \\
17 & 2.00 & 483.10 & NS \\
18 & 4.26 & 102.03 & NS \\
18.50 & 3.91 & 270.07 & NS \\
19 & 4.04 & 173.01 & NS \\
19.56 & 3.97 & 433.22 & \textbf{BH} \\
20 & 3.89 & 250.09 & NS \\
21.68 & 2.07 & 289.02 & NS \\
23 & 4.19 & 250.17 & NS \\
24 & 3.88 & 265.04 & NS \\
25 & 3.89 & 277.30 & NS \\
40 & 1.76 & 936.01 & \textbf{BH} \\
60 & 4.49 & 216.07 & NS \\
\enddata
\tablecomments{This table summarizes GW waveform models from \citet{2023PhRvD.107j3015V, 2024arXiv240801525C}, with progenitor mass, simulation duration, GW energy emitted ($\mathrm{E_{\rm GW}}$) through matter, and compact object outcome (NS or BH). Models denoted by an asterisk (*) fail to explode and directly collapse into black holes. In contrast, models such as the 19.56 M$_\odot$ and 40 M$_\odot$ cases initially explode before forming BHs \citep{2023ApJ...957...68B}. The SFHo equation of state and the M1 neutrino transport scheme \citep{O_Connor_2018} are used throughout. Radiated energy spans three orders of magnitude, with general trends of increasing energy with progenitor mass. Deviations from monotonicity highlight the diversity in the dynamics of the CCSN.}
\end{deluxetable}

\clearpage
\subsection{Establishing a Detection Metric}

The detectability of GWs from CCSNe is quantified using the signal-to-noise ratio (SNR), defined as:

\begin{equation} 
{\rm SNR} = \sqrt{4 \int_{f_{\min}}^{\infty} \frac{|\tilde{h}(f)|^{2}}{S_{n}(f)} \, df}, 
\end{equation}

where $\tilde{h}(f)$ is the GW strain in the frequency domain, \(S_n(f)\) is the detector noise power spectral density (PSD) and \(f_{\min}\) is the minimum frequency of the GW signal \citep{2013arXiv1304.0210S}. 

For this study, the CCSNe GW waveforms are scaled to a fiducial distance of 10 kpc and evaluated against the sensitivity curves of current terrestrial detectors (e.g., LIGO, A$\#$, CE, ET) and proposed future space-based and lunar observatories (e.g., B-DECIGO, DECIGO, TianGO, DO, ALIA, Atomic Clock, GLOC, LILA, LGWA). This analysis serves as a benchmark for evaluating the detection capabilities of various GW detectors across a range of progenitor models. 

\subsection{Extending GW Models}

The CCSNe models capture several seconds of post-bounce evolution but do not resolve GW signal evolution below 0.1 Hz due to current computational constraints. To address this limitation, we extend the models using the tapering technique of \citet{2022PhRvD.105j3008R}, which analytically models the long-term evolution of neutrino-driven anisotropies. This extension allows us to probe GW signals into the sub-Hz regime.

For this study, we applied a cosine-modulated tapering function from \citet{2022PhRvD.105j3008R} to smoothly reduce the GW strain to zero over a 10-second interval, minimizing artifacts caused by abrupt signal termination:

\begin{equation}
h_{\times /+}^{\text{tail}}=\frac{h_{\times /+}^{\text{end}}}{2}\left[1+\cos \left(2 \pi f_{t}\left(t-t^{\text{end}}\right)\right)\right],
\end{equation}

where \(f_t\) is the frequency of the tapering function, \(t^{\text{end}}\) is the simulation's truncation time, and \(h^{\text{end}}\) is the GW strain at \(t^{\text{end}}\). This approach ensures smooth attenuation of the signal, avoiding discontinuities that could introduce artifacts in the frequency domain. However, tapering may suppress low-frequency components, leading to conservative SNR estimates. For further explanation, please see the Appendix.

An alternative method to extend GW signals is the Linear Prediction Filter (LPF) \citep{2024arXiv240402131R}. Although this approach has shown initial promise in preserving low-frequency secular characteristics, improper application of LPF can distort signals and artificially inflate SNRs, particularly at frequencies below 200 Hz. For this reason, we adopted the tapering function to ensure robust and conservative results.

To further explore the impact of tapering on the SNR estimates, we conducted a supplementary test in which the truncated GW signal was extended using the actual neutrino luminosity data correlated with the GW models. This approach preserves the physical integrity of the signal while addressing the dampening effect introduced by the tapering. As detailed in the Appendix, the comparison highlights the differences in SNR outcomes between extension methods.


\section{Gravitational Wave Detection Prospects}\label{sec:res}

In this section, we examine the detection prospects for low-frequency GWs from CCSNe, emphasizing the transformative potential of future space-based and lunar detectors. By evaluating the modeled GW strain across a diverse range of CCSNe progenitors and comparing these signals to the sensitivity curves of current and next-generation detectors, we identify the instruments best suited to capture both matter-sourced and neutrino-sourced GW emissions. Central to this analysis is the GW linear memory effect, a persistent low-frequency signal that holds the promise of significantly enhancing the detectability of CCSNe GWs. Harnessing this feature may enable breakthroughs in detection rates, extending beyond the limits of existing terrestrial observatories and offering new insights into the physics of stellar collapse.

Figures \ref{fig:neutrino} and \ref{fig:matter} present the amplitude spectral densities (ASDs) of GW generated from matter and neutrino emission in CCSNe for progenitors with masses ranging from 9 to 60 M$_\odot$ at a distance of 10 kpc. These spectra are compared with the noise curves \(\sqrt{S_n}\) of various GW detectors, including terrestrial facilities (LIGO O3b, A$\#$, CE, and ET), proposed deci-hertz observatories (B-DECIGO, DECIGO and Deci-Hz Observatory) and future space and lunar detectors (TianGO, BBO, ALIA, Atomic Clock, GLOC, LILA, and LGWA). By evaluating this broad suite of detectors, we identify which instruments are most capable of capturing both matter- and neutrino-sourced GW signals, providing a comprehensive perspective on detection strategies across frequency regimes.

For neutrino-sourced GW signals (Figure \ref{fig:neutrino}), a distinctive low-frequency memory effect below 1 Hz arises from anisotropic neutrino emission during the collapse phase. This long-duration, nearly constant offset in GW strain significantly enhances signal strengths compared to matter-sourced emissions. Lunar and deci-hertz detectors are particularly well suited to capture these low-frequency signatures, providing critical insights into the dynamics of neutrino-driven processes. As summarized in Table~\ref{tab:neutrinospace}, DECIGO achieves exceptional sensitivity, with total SNRs exceeding $10^4$ for high-mass progenitors like 19.56 M$_\odot$, while GLOC achieves over $10^3$ for the same progenitor at 10 kpc. This makes these detectors indispensable for studying low-frequency neutrino-driven GWs at both galactic and extra-galactic scales.

In contrast, matter-sourced GW signals (Figure \ref{fig:matter}) feature low-frequency components below 100 Hz that mainly reflect large-scale mass motions during the collapse phase. These signals are optimally observed with both deci-Hz and lunar-based detectors, which excel in this frequency range. Frequencies above 100 Hz, associated with PNS formation and dynamics, are accessible to terrestrial observatories like CE and ET, which are designed to operate effectively in the high-frequency regime.

The performance of GW detectors across different progenitor masses is summarized in Tables~\ref{tab:matter_snr_space_lunar_updated} - ~\ref{tab:neutrinospace}. Space-based observatories, such as DECIGO and BBO, consistently achieve exceptional sensitivity to low-frequency signals, with total SNRs exceeding 600 for matter-sourced emissions and surpassing $10^4$ for neutrino-driven signals, enabling detection horizons beyond 10 Mpc. Similarly, advanced lunar detectors such as GLOC demonstrate strong sensitivity, achieving an SNR of 660.49 for the 19.56 M$_\odot$ progenitor at 10 kpc, with significant detectability extending up to 10 Mpc. These observatories provide invaluable tools for studying low-frequency features, such as the GW linear memory effect, explosion asymmetries, and compact object formation.

In contrast, terrestrial detectors like LIGO O3b face limitations in detecting low-frequency signals due to seismic noise constraints, with SNRs often falling below 1 for low-mass progenitors at Galactic distances. However, next-generation detectors such as CE and ET exhibit substantial improvements, achieving SNRs exceeding 300 for neutrino-sourced signals from Galactic CCSNe, particularly in the high-frequency regime (100–1000 Hz) where features like PNS formation dominate. These detectors remain essential for probing signals from Galactic CCSNe but are less equipped to capture the low-frequency dynamics of stellar collapse.

Lunar-based observatories, particularly advanced designs like GLOC, bridge the gap between terrestrial- and space-based detectors.~With their unique sensitivity to low-frequency signals, they complement space-based observatories like DECIGO by offering a scalable solution for probing CCSNe within 10 Mpc and beyond. First-generation lunar detectors, such as LILA and LGWA, though more limited in capability, represent critical stepping stones in developing this frontier, enabling Galactic-scale detections and paving the way for more advanced designs.

Across the spectrum, the mass of the progenitor plays a critical role in shaping GW emission. High-mass progenitors (for example, 40-60 M$_\odot$) generate amplitudes of up to two orders of magnitude stronger than lower-mass counterparts (for example, 9 M$_\odot$), with advanced detectors like DECIGO, GLOC, and ET offering the sensitivity required to capture these signals. These complementary detector platforms collectively expand our understanding of CCSNe, from local events within the Milky Way to extragalactic sources, offering a comprehensive view of stellar collapse and gravitational wave emission mechanisms.

Across the spectrum of current and future detectors, space-based and lunar observatories stand out as uniquely equipped to probe the low-frequency regime, where the GW linear memory effect dominates. Terrestrial detectors, such as CE and ET, excel in detecting high-frequency signals associated with PNS formation but are limited by seismic noise in the deci-hertz range. Space-based observatories like DECIGO and BBO achieve unparalleled sensitivity to both matter- and neutrino-sourced gravitational waves, extending detection horizons far beyond 10 Mpc. Advanced lunar designs, such as GLOC, complement these efforts with superior low-frequency capabilities, effectively bridging the gap between terrestrial and space-based observatories.

\section{Conclusions}\label{sec:con}
The detection of GWs from CCSNe presents an unparalleled opportunity to explore the intricate dynamics of stellar collapse and explosion, as well as the formation of compact objects. This study evaluates the feasibility of detecting both matter-sourced and neutrino-sourced GW emission from CCSNe for a range of current and future GW detectors, with a special focus on lunar- and space-based GW observatories. Several key findings emerge from this analysis, providing a pathway for future detection strategies and a deeper understanding of stellar death and the compact remnants they leave behind.

First, we reaffirm the limitations of current terrestrial detectors, such as LIGO, in capturing GWs from lower-mass CCSNe progenitors. The sensitivity of LIGO is ultimately reliably restricted to higher frequencies (100-1000 Hz), limiting its ability to detect GWs to the nearby Galactic CCSNe. However, next-generation detectors such as CE and ET show immense promise. Their enhanced sensitivity across 10-1000 Hz suggests that they will be capable of detecting GWs from more massive progenitors, even from extragalactic distances, offering insight into diverse stellar collapse scenarios.

Second, lunar- and space-based detectors—particularly GLOC, LILA, and DECIGO—are poised to transform our understanding of low-frequency GW emissions. These detectors are optimally suited for detecting the GW linear memory effect, a low-frequency signal associated with asymmetric neutrino emissions during stellar collapse. By detecting this subtle yet crucial signal, lunar and space-based observatories can significantly extend our GW detection horizon to several megaparsecs, probing CCSNe events in nearby galaxies. This aligns with conclusions from previous studies, such as \cite{2024MNRAS.532.4326P} and \cite{2024arXiv240801525C}, which similarly emphasize the critical role of deci-hertz and lunar detectors in detecting GWs from CCSNe.

The detection of the GW linear memory effect would mark a monumental achievement in GW astronomy, offering direct confirmation of one of the key predictions of general relativity. The memory effect, driven by the asymmetric ejection of matter and neutrinos, carries essential information about the internal dynamics of the explosion and the formation of compact objects. Beyond confirming general relativity, detecting this effect would provide a new window into understanding the asymmetric processes that drive these events, allowing us to probe the conditions inside the collapsing star with unprecedented precision.


Moreover, CCSNe associated with long gamma-ray bursts (GRBs) offer a complementary avenue for exploration. Rapidly rotating progenitors collapsing into black holes are believed to produce GRBs, which emit their own GW memory signals from the anisotropic ejection of matter and energy \citep{2013PhRvD..87l3007B, 2013PASJ...65...59A}. Detecting these off-axis GW signals from GRBs, especially at high redshifts where electromagnetic signals may be faint or obscured, could provide critical insights into extreme cosmic events. Future space-based detectors like LISA and DECIGO are ideally suited to detect these signals, offering a glimpse into the violent processes shaping the early universe \citep{2023MNRAS.518.5242U}.

In summary, the combination of terrestrial, lunar, and space-based observatories offers an exciting path forward to detect GWs from CCSNe and GRBs. Terrestrial detectors, both current and next generation, will continue to provide insights into the higher-frequency signals from massive progenitors, while lunar and space-based detectors will unlock the low-frequency GW domain, revealing the GW linear memory effect, neutrino-sourced GWs, and GRB-related signals. Together, these advancements will not only confirm key predictions of general relativity but also open new frontiers in understanding the most extreme events in the universe: from stellar collapse and compact object formation to the emission of long GRBs in the early universe.

Looking ahead, the pursuit of detecting low-frequency GWs from CCSNe and GRBs will benefit immensely from innovative approaches that combine advanced machine learning techniques with next-generation observational platforms. This study underscores the transformative potential of lunar-based GW observatories in probing sub-Hz to Hz frequency signals, including the elusive GW linear memory effect. However, unlocking this potential requires new methods capable of operating within the unique lunar noise environment.

To address these challenges, our follow-up work introduces the first convolutional neural network combined with a long short-term memory network, specifically designed to detect GW emissions from both compact binary mergers and massive stellar collapses in simulated lunar noise. This groundbreaking framework not only extends the application of machine learning in GW detection but also lays the foundation for exploring the asymmetric dynamics of CCSNe and the formation of compact objects with unparalleled precision. By enabling detection of both merging binaries and stellar collapse events, this approach bridges the gap between theory and observation, opening an entirely new window into the sub-Hz GW spectrum.

As we advance our tools and infrastructure for lunar GW detection, the synergy between cutting-edge machine learning frameworks and innovative detector design will drive the next era of GW astronomy. The quest to capture the faint whispers of stellar death and dynamic cosmic events has just begun, and the forthcoming insights promise to transform our understanding of the universe’s most extreme phenomena.

\begin{deluxetable*}{c|cccc|c|cccc}
\setlength{\tabcolsep}{10pt}
\tablewidth{0pt}
\tablecaption{\textbf{Total and Partial SNR for Matter-Sourced and Neutrino-Sourced GW Strain (Terrestrial Observatories)}\label{tab:optimized1}}
\tablehead{
    \colhead{Progenitor Model (M$_\odot$)} & \colhead{LIGO O3} & \colhead{A$\#$} & \colhead{CE} & \colhead{ET} &
    \colhead{Progenitor Model (M$_\odot$)} & \colhead{LIGO O3} & \colhead{A$\#$} & \colhead{CE} & \colhead{ET}
}
\startdata
\tableline \hline
\multicolumn{5}{c|}{\textbf{Matter-Sourced Total SNR}} & \multicolumn{5}{c}{\textbf{Neutrino-Sourced Total SNR}} \\ \hline
9a    & 0.93  & 5.9  & 15   & 16   & 9a    & 0.90  & 3.2  & 33   & 35   \\ \hline
9b    & 1.4   & 8.2  & 22   & 23   & 9b    & 0.59  & 4.1  & 41   & 29   \\ \hline
9.25  & 1.7   & 12   & 31   & 30   & 9.25  & 1.5   & 8.9  & 70   & 63   \\ \hline
9.50  & 2.8   & 18   & 48   & 47   & 9.50  & 2.2   & 8.6  & 75   & 94   \\ \hline
11    & 4.7   & 33   & 88   & 81   & 11    & 38    & 13   & 120  & 100  \\ \hline
12.25 & 2.7   & 20   & 52   & 49   & 12.25 & 9.4   & 23   & 150  & 120  \\ \hline
14    & 4.8   & 30   & 83   & 79   & 14    & 7.3   & 37   & 250  & 200  \\ \hline
15.01 & 5.6   & 43   & 112  & 101  & 15.01 & 12    & 23   & 200  & 220  \\ \hline
16    & 8.9   & 54   & 163  & 144  & 16    & 16    & 18   & 140  & 130  \\ \hline
17    & 13    & 81   & 235  & 209  & 17    & 21    & 27   & 230  & 260  \\ \hline
18    & 11    & 68   & 203  & 179  & 18    & 8.4   & 22   & 170  & 170  \\ \hline
18.5  & 11    & 68   & 205  & 181  & 18.5  & 30    & 40   & 360  & 310  \\ \hline
19    & 11    & 69   & 202  & 179  & 19    & 49    & 21   & 170  & 160  \\ \hline
19.56 & 16    & \textbf{102}  & \textbf{301}  & \textbf{263}  & 19.56 & 27    & \textbf{44}   & \textbf{380}  & 330  \\ \hline
20    & 11    & 67   & 200  & 178  & 20    & 8.3   & 41   & 350  & 330  \\ \hline
21.68 & 11    & 72   & 202  & 182  & 21.68 & 23    & 34   & 240  & 240  \\ \hline
23    & 6.6   & 44   & 122  & 112  & 23    & 17    & 27   & 250  & 220  \\ \hline
24    & 11    & 69   & 205  & 181  & 24    & 20    & 32   & 290  & 270  \\ \hline
25    & 12    & 74   & 219  & 194  & 25    & 16    & 39   & 350  & 340  \\ \hline
40    & \textbf{15}    & 96   & 278  & 243  & 40    & 17    & 40   & 330  & \textbf{360}  \\ \hline
60    & 10    & 67   & 195  & 171  & 60    & \textbf{40}    & 26   & 230  & 230  \\ 
\tableline \hline
\multicolumn{5}{c|}{\textbf{Matter-Sourced Partial SNR (0.1 - 10 Hz)}} & \multicolumn{5}{c}{\textbf{Neutrino-Sourced Partial SNR (0.1 - 10 Hz)}} \\ \hline
9a    & -     & -    & -    & 0.37 & 9a    & 0.83  & 0.34 & 4.3  & 30   \\ \hline
9b    & -     & -    & -    & 0.46 & 9b    & 0.57  & 0.16 & 2.2  & 20   \\ \hline
9.25  & -     & -    & 0.12 & 0.59 & 9.25  & 1.4   & 0.51 & 8.0  & 48   \\ \hline
9.50  & -     & -    & 0.22 & 1.0  & 9.50  & 2.0   & 0.80 & 12   & 83   \\ \hline
11    & -     & -    & 0.26 & 1.5  & 11    & 26    & 0.77 & 10   & 73   \\ \hline
12.25 & -     & -    & -    & 0.11 & 12.25 & 6.9   & 1.1  & 17   & 73   \\ \hline
14    & -     & -    & -    & 0.87 & 14    & 4.1   & 1.6  & 23   & 140  \\ \hline
15.01 & -     & -    & 0.57 & 3.4  & 15.01 & 9.8   & 2.3  & 31   & 200  \\ \hline
16    & -     & -    & 0.76 & 4.1  & 16    & 7.9   & 1.2  & 16   & 100  \\ \hline
17    & -     & -    & 0.98 & 4.9  & 17    & 19    & 2.6  & 35   & 220  \\ \hline
18    & -     & -    & 1.1  & 5.2  & 18    & 7.6   & 1.7  & 29   & 130  \\ \hline
18.5  & -     & -    & 1.6  & 7.4  & 18.5  & 28    & 2.4  & 28   & 240  \\ \hline
19    & -     & -    & 1.7  & 6.3  & 19    & \textbf{34}    & 1.3  & 20   & 130  \\ \hline
19.56 & -     & 0.24 & 3.6  & 15   & 19.56 & 20    & 2.9  & 44   & 250  \\ \hline
20    & -     & -    & 0.74 & 4.5  & 20    & 7.5   & \textbf{3.5}  & \textbf{49}   & 270  \\ \hline
21.68 & -     & -    & 1.4  & 7.2  & 21.68 & 19    & 1.9  & 27   & 190  \\ \hline
23    & -     & -    & 0.67 & 4.0  & 23    & 15    & 2.2  & 33   & 170  \\ \hline
24    & -     & -    & 1.6  & 8.4  & 24    & 17    & 2.1  & 31   & 200  \\ \hline
25    & -     & -    & 1.1  & 8.2  & 25    & 14    & 2.8  & 34   & 270  \\ \hline
40    & -     & 0.27 & \textbf{4.5}  & \textbf{16}   & 40    & 16    & 2.2  & 28   & \textbf{300}  \\ \hline
60    & -     & -    & 0.78 & 3.7  & 60    & 28    & 1.8  & 27   & 190  \\ 
\tableline \hline
\enddata
\tablecomments{ 
This table summarizes the total and partial SNRs for GWs generated by matter dynamics and neutrino-driven anisotropies in CCSNe. Calculated for progenitor models at 10 kpc, it evaluates four GW observatories: LIGO O3, A$\#$, Cosmic Explorer (CE), and Einstein Telescope (ET). Total SNR columns capture GW detectability across the full frequency spectrum, while Partial SNR columns (0.1–10 Hz) highlight low-frequency components critical for features like GW linear memory and neutrino anisotropy emissions. Left and right sections compare matter- and neutrino-sourced strains. Bolded SNR values indicate the progenitor mass with the highest SNR across detector configurations, while dashes ("-") denote negligible SNR. Results highlight the sensitivity gains of next-generation detectors, especially in the low-frequency regime.
}
\end{deluxetable*}

\samepage 
\begin{deluxetable*}{c|ccccccc|ccc}[h!]
\setlength{\tabcolsep}{5pt}
\tablewidth{0pt}
\tablecaption{\textbf{Matter-Sourced Total SNR for Space-based and Lunar GW Observatories}}
\label{tab:matter_snr_space_lunar_updated}
\tabletypesize{\scriptsize}
\tablehead{
    \colhead{Progenitor Model (M$_\odot$)} & \colhead{DECIGO} & \colhead{B-DECIGO} & \colhead{BBO} & \colhead{DO} & \colhead{ALIA} & \colhead{AC} & \colhead{TianGO} & \colhead{GLOC} & \colhead{LILA} & \colhead{LGWA}
}
\startdata
9a & 11.83 & 0.52 & 2.98 & 2.10 & 0.51 & 0.14 & 0.07 & 31.75 & 0.15 & 0.02 \\
9b & 1.50 & 0.11 & 0.64 & 0.24 & 0.04 & 0.01 & 0.02 & 45.85 & 0.17 & 0.00 \\
9.25 & 6.57 & 0.56 & 3.41 & 1.08 & 0.20 & 0.05 & 0.06 & 63.18 & 0.41 & 0.02 \\
9.5 & 49.37 & 2.05 & 11.57 & 8.79 & 2.18 & 0.58 & 0.30 & 100.50 & 0.46 & 0.08 \\
11 & 110.79 & 3.78 & 18.62 & 20.32 & 5.48 & 1.45 & 0.52 & 181.11 & 1.53 & 0.14 \\
12.25 & 0.55 & 0.02 & 0.62 & 0.10 & 0.03 & 0.01 & 0.02 & 109.94 & 0.07 & 0.00 \\
14 & 9.42 & 0.51 & 3.01 & 1.64 & 0.39 & 0.10 & 0.10 & 190.73 & 0.37 & 0.02 \\
15.01 & 123.58 & 4.45 & 22.77 & 22.60 & 6.06 & 1.60 & 0.63 & 247.86 & 1.83 & 0.17 \\
16 & 74.16 & 4.18 & 24.88 & 12.68 & 2.76 & 0.73 & 0.56 & 352.34 & 1.43 & 0.16 \\
17 & 570.76 & 25.30 & 144.58 & 101.03 & 24.64 & 6.51 & 3.55 & 534.08 & 3.99 & 0.98 \\
18 & 119.93 & 4.83 & 26.14 & 21.71 & 5.68 & 1.50 & 0.63 & 434.65 & 2.28 & 0.18 \\
18.5 & 108.50 & 4.82 & 26.77 & 19.71 & 5.23 & 1.38 & 0.56 & 444.27 & 4.29 & 0.16 \\
19 & 190.20 & 7.41 & 38.69 & 34.71 & 9.27 & 2.45 & 0.93 & 435.76 & 3.16 & 0.27 \\
19.56 & \textbf{621.97} & \textbf{27.01} & \textbf{149.88} & \textbf{111.53} & \textbf{28.48} & \textbf{7.52} & \textbf{3.78} & \textbf{660.49} & 7.60 & \textbf{1.03} \\
20 & 50.22 & 2.32 & 13.10 & 9.02 & 2.32 & 0.61 & 0.26 & 447.86 & 2.50 & 0.08 \\
21.68 & 229.47 & 11.85 & 69.49 & 40.05 & 9.36 & 2.47 & 1.50 & 424.60 & 4.14 & 0.44 \\
23 & 62.66 & 2.93 & 17.53 & 10.87 & 2.46 & 0.65 & 0.41 & 258.44 & 1.86 & 0.11 \\
24 & 593.58 & 24.08 & 131.95 & 106.85 & 27.45 & 7.25 & 3.35 & 455.67 & 4.07 & 0.92 \\
25 & 116.45 & 8.42 & 50.56 & 19.71 & 4.21 & 1.11 & 0.78 & 487.91 & 4.10 & 0.27 \\
40 & 451.00 & 25.99 & 154.04 & 77.50 & 17.21 & 4.55 & 3.14 & 606.85 & \textbf{8.64} & 0.94 \\
60 & 89.86 & 4.60 & 26.38 & 15.99 & 4.02 & 1.06 & 0.57 & 429.95 & 2.35 & 0.16 \\
\enddata
\tablecomments{
    Total SNRs for matter-sourced strains at 10 kpc are presented for space-based and lunar detectors: DECIGO, B-DECIGO, BBO, DO, ALIA, Atomic Clock (AC), TianGO, GLOC, LILA, and LGWA. The vertical line between TianGO and GLOC indicates the division between space-based and lunar GW detectors. The columns are ordered by detectors with decreasing SNR, with DECIGO presenting the highest SNR. The progenitor model with the consistently highest SNR, 19.56 M\(_\odot\), is boldfaced across all detectors, except for LILA, where the 40 \(M_\odot\) progenitor produces the highest SNR. 
}
\end{deluxetable*}

\begin{deluxetable*}{c|ccccccc|ccc}[h!]
\setlength{\tabcolsep}{5pt}
\tablewidth{0pt}
\tablecaption{\textbf{Neutrino-Sourced Total SNR for Space-based and Lunar GW Observatories}\label{tab:neutrinospace}}
\tabletypesize{\scriptsize}
\tablehead{
    \colhead{Progenitor Model (M$_\odot$)} & \colhead{DECIGO} & \colhead{B-DECIGO} & \colhead{BBO} & \colhead{DO} & \colhead{ALIA} & \colhead{AC} & \colhead{TianGO} & \colhead{GLOC} & \colhead{LILA} & \colhead{LGWA}
}
\startdata
9a    & 417    & 24.9   & 149    & 71.1   & 15.3   & 4.03   & 3.03   & 118    & 14.4   & 0.90   \\ 
9b    & 249.63 & 16.53  & 99.72  & 41.38  & 7.94   & 2.10   & 1.99   & 124.22 & 13.77  & 0.59   \\ 
9.25  & 651.75 & 41.48  & 253.58 & 108.76 & 20.90  & 5.53   & 4.93   & 226.69 & 25.18  & 1.47   \\ 
9.50  & 963.36 & 61.05  & 366.65 & 162.25 & 33.13  & 8.76   & 7.29   & 352.76 & 46.58  & 2.18   \\ 
11    & 25354.25 & 961.49 & 5202.83 & 4570.26 & 1175.24 & 310.35 & 142.10 & 412.91 & 66.19  & 37.98  \\ 
12.25 & 6122.26 & 238.01 & 1315.19 & 1097.36 & 277.32  & 73.23  & 35.10  & 433.44 & 47.85  & 9.41   \\ 
14    & 5314.67 & 194.71 & 1021.17 & 968.59  & 256.76  & 67.79  & 25.79  & 742.48 & 119.72 & 7.27   \\ 
15.01 & 6835.87 & 303.25 & 1785.86 & 1194.11 & 275.06  & 72.70  & 39.85  & 903.31 & 140.18 & 11.64  \\ 
16    & 12210.66 & 436.70 & 2163.47 & 2246.24 & 611.72  & 161.49 & 57.43  & 517.33 & 97.98  & 16.33  \\ 
17    & 8887.32 & 574.65 & 3460.58 & 1490.44 & 298.21  & 78.85  & 68.33  & 919.07 & 183.95 & 20.63  \\ 
18    & 4083.61 & 239.38 & 1408.01 & 703.90  & 157.75  & 41.69  & 23.90  & 671.80 & 63.77  & 8.40   \\ 
18.5  & 14786.21 & 769.39 & 4739.20 & 2483.70 & 479.54  & 126.89 & 115.20 & 1185.21 & 92.16  & 30.33  \\ 
19    & \textbf{31543.49} & \textbf{1259.12} & \textbf{6807.44} & \textbf{5697.79} & \textbf{1479.18} & \textbf{390.58} & \textbf{177.52} & 669.05 & 128.97 & \textbf{48.50} \\ 
19.56 & 17820.10 & 705.41 & 3947.76 & 3184.65 & 796.11  & 210.29 & 100.72 & 1338.51 & 155.26 & 27.46  \\ 
20    & 4195.10 & 245.78 & 1501.11 & 717.80  & 154.59  & 40.87  & 27.95  & 1185.37 & 139.60 & 8.30   \\ 
21.68 & 12783.57 & 618.06 & 3593.77 & 2253.93 & 544.51  & 143.82 & 79.13  & 991.34  & 218.12 & 22.71  \\ 
23    & 8670.21 & 431.33 & 2616.37 & 1479.37 & 311.08  & 82.24  & 64.66  & 870.83  & 149.90 & 17.01  \\ 
24    & 10931.95 & 502.47 & 2913.13 & 1921.89 & 459.38  & 121.35 & 74.86  & 1066.75 & 125.99 & 19.69  \\ 
25    & 7435.61 & 441.44 & 2601.83 & 1290.77 & 301.90  & 79.72  & 53.37  & 1378.25 & 257.46 & 15.70  \\ 
40    & 8177.16 & 480.48 & 2905.21 & 1396.64 & 301.63  & 79.72  & 59.24  & \textbf{1400.32} & \textbf{281} & 17.13  \\ 
60    & 24937.96 & 1074.89 & 5812.54 & 4520.84 & 1191.93 & 314.65 & 140.28 & 996.08  & 186.61 & 39.75  \\ 
\enddata
\tablecomments{
    Total SNRs for neutrino-sourced strains at 10 kpc are presented for space-based and lunar detectors: DECIGO, B-DECIGO, BBO, DO, ALIA, Atomic Clock (AC), TianGO, GLOC, LILA, and LGWA. The vertical line between TianGO and GLOC separates space-based from lunar detectors.  The columns are ordered by detectors with decreasing SNR, with DECIGO presenting the highest SNR. The progenitor model with the consistently highest SNR, 19 \(M_\odot\), is boldfaced across all detectors, except for GLOC and LILA, where the 40 \(M_\odot\) progenitor produces the highest SNR.
}
\end{deluxetable*}

\bibliography{sample62}


\section*{Appendix}

\begin{center}
    \textbf{-- Searching for Light Knights in the Dark -- \\ An Ode to the Oppenheimer Crater}
\end{center}

\begin{center}
    Red rubies afar, from galaxies wide, \\ 
    Through spacetime, they twist and collide. \\ 
    The Black Hole Moon, sentinel high, silent and still, \\ 
    Catches their collapse, bends it to our will. \\ 
    Here stars meet their fate, in gravitational song, \\ 
    While the universe’s birth echoes strong. \\ 
    To unlock these secrets, we must place a key, \\ 
    On the Moon’s dark expanse, where truth breaks free. \\ 
    The waves of their dying, the whispers of old, \\ 
    The stories of stars and spacetime retold. \\ 
\end{center}

\section*{Acknowledgements}
The author extends heartfelt gratitude to Josh Grindlay, David Vartanyan, Martin Elvis, Peter Fritschel, Lisa Kewley, Kim Vy-Tran and John A. Lewis for their invaluable contributions through creative discussions and meticulous code review. Special thanks are also owed to Avi Loeb, Karan Jani, Marica Branchesi, and Jan Harms for their insightful guidance and inspiring ideas, which greatly enriched this work. The author lastly thanks Adam Burrows for his extensive, publicly available CCSNe GW models and for simply being \texttt{cool}. This research has made use of data or software obtained from the Gravitational Wave Open Science Center (gw-openscience.org), a service of LIGO Laboratory, the LIGO Scientific Collaboration, the Virgo Collaboration, and KAGRA. LIGO Laboratory and Advanced LIGO are funded by the United States National Science Foundation (NSF) as well as the Science and Technology Facilities Council (STFC) of the United Kingdom, the Max-Planck-Society (MPS), and the State of Niedersachsen/Germany for support of the construction of Advanced LIGO and construction and operation of the GEO600 detector. Additional support for Advanced LIGO was provided by the Australian Research Council. Virgo is funded, through the European Gravitational Observatory (EGO), by the French Centre National de Recherche Scientifique (CNRS), the Italian Istituto Nazionale di Fisica Nucleare (INFN) and the Dutch Nikhef, with contributions by institutions from Belgium, Germany, Greece, Hungary, Ireland, Japan, Monaco, Poland, Portugal, Spain. The construction and operation of KAGRA are funded by Ministry of Education, Culture, Sports, Science and Technology (MEXT), and Japan Society for the Promotion of Science (JSPS), National Research Foundation (NRF) and Ministry of Science and ICT (MSIT) in Korea, Academia Sinica (AS) and the Ministry of Science and Technology (MoST) in Taiwan. All affiliated code is publicly available at \texttt{https://github.com/kiranjyot/MoonMemory}. 
\end{document}